\documentstyle[aps,twocolumn,epsfig]{revtex}
\begin{document}
\newcommand{\ve}[1]{\mbox{\boldmath $#1$}}
\twocolumn[\hsize\textwidth\columnwidth\hsize
\csname@twocolumnfalse%
\endcsname

\draft

\title{Effectively attractive Bose-Einstein condensates in a 
rotating toroidal trap} 
\author{G.~M.~Kavoulakis} 
\date{\today} 
\address{Mathematical Physics, Lund Institute of Technology, P.O.  Box 
118, S-22100 Lund, Sweden}
	
\maketitle

\begin{abstract}

We examine an effectively attractive quasi-one-dimensional Bose-Einstein
condensate of atoms confined in a rotating toroidal trap, as the magnitude
of the coupling constant and the rotational frequency are varied. 
Using both a variational mean-field approach, as well as a diagonalization
technique, we identify the phase diagram between a uniform and a localized 
state and we describe the system in the two phases.

\end{abstract}

\vskip0.0pc]

Effectively one-dimensional clouds of trapped atoms at very low 
temperatures are interesting systems because of a number of reasons.  
First of all, highly anisotropic traps allow us to realize this 
situation experimentally \cite{Kettnew}.  For example, in elongated 
traps, when the oscillator energy transversely to the long axis of the 
trap is much larger than the strength of the interaction, the degrees 
of freedom along this direction are frozen out and the system is 
effectively one-dimensional.  In addition, such systems are 
interesting from a theoretical point of view since they allow us to 
check one-dimensional models which have been developed over the years.

Recently Strecker {\it et al.} \cite{sol1} and Khaykovich {\it et al.} 
\cite{sol2} have managed to create and observe bright solitons -- 
localized blobs of atoms -- in quasi-one-dimensional clouds of atoms after 
switching the effective coupling constant from positive (corresponding 
to an effective repulsive interaction between the atoms) to negative 
(corresponding to an effective attractive interaction) with use of the 
so-called Feshbach resonances \cite{Ket}.  References \cite{refs} have 
examined these experiments theoretically.

Inspired by these experiments, Refs.\,\cite{Ueda} and \cite{GK} have 
studied the behavior of a cloud of atoms confined in a toroidal 
trap.  Their basic conslusion is that while the density of such a 
system is homogeneous for an effective repulsion between the atoms, 
when the interaction becomes attractive, there is a critical value of 
the coupling constant below which the atoms form a localized blob.  An 
easy way of understanding this effect is that the kinetic energy that 
results from the Heisenberg uncertainty principle favors a homogeneous 
density, whereas an attractive interaction between the atoms favors 
the formation of a localized density.

Extending their study of Ref.\,\cite{Ueda}, in a more recent paper 
Kanamoto {\it et al.} have examined the same problem when the toroidal 
trap is rotated \cite{Ueda2}. While in a non-rotating torus there is 
only one parameter to vary, i.e, the ratio between the interaction 
and the kinetic energy (which we call $\gamma$), in the present 
problem there is an extra degree of freedom, namely the rotational 
frequency of the torus $\Omega$.  Using the mean-field approximation, 
as well as numerical diagonalization of the Hamiltonian, Kanamoto 
{\it et al.} have identified the phase diagram that separates the 
phase of uniform density from the phase of localized density in the 
$\Omega$ -- $\gamma$ plane.  

In our study we attack the same problem using different methods, and 
our results are in agreement with those of Ref.\,\cite{Ueda2}.  In the 
first method we use the mean-field approximation and calculate the 
order parameter variationally.  In the second method we truncate 
appropriately the Hamiltonian of the system and diagonalize it 
analytically using a Bogoliubov tranformation.  Our results are also 
relevant with those of Lundh {\it et al.} \cite{Emil}, where the 
rotational properties of an attractive Bose gas that is confined in an 
anharmonic potential have been investigated.  As shown in this study, 
the effective potential felt by the atoms is toroidal and
the angular momentum is carried either by exciting the center of mass, 
or by creating vortex states.  One may argue that the localized phase 
of our problem corresponds to the center of mass excitation, while the 
uniform phase corresponds to the vortex phase \cite{pr}.

Let us therefore consider a Bose-Einstein condensate of $N$ atoms confined  
in a toroidal trap which has a radius $R$ and a cross section $S=\pi r^{2}$, 
where we assume that $r \ll R$.  Measuring the length in units of $R$ and 
the energy in units of $\hbar^2/2 M R^2$, with $M$ being the atom 
mass, the Hamiltonian $\hat{H}$ in the rotating frame of reference is
\begin{eqnarray}
\hat{H} = \int_0^{2\pi} d\theta
       \left[\hat{\psi}^{\dagger}(\theta) ({\hat L} - \Omega)^{2} 
       \hat{\psi}(\theta) \right.
\nonumber \\ \left.
       +\frac 1 2 U_0 \hat{\psi}^{\dagger}(\theta)\hat{\psi}^{\dagger}(\theta) 
       \hat{\psi}(\theta)\hat{\psi}(\theta)\right].
\label{Ham}
\end{eqnarray}
Here $\hat{\psi}(\theta)$ is the field operator, ${\hat L} = -i 
\partial/\partial \theta$ is the operator of the angular momentum, 
with $\theta$ being the azimuthal angle, $2 \Omega$ is the angular 
frequency of rotation, and $U_0 = 8\pi aR/S$, with $a$ being the 
scattering length for elastic atom-atom collisions. In the above 
expression the energy has been shifted by the term $\Omega^{2}$ for 
convenience.

Let us start with the mean-field approximation.  Within this scheme 
the system is described by the order parameter ${\psi}(\theta)$.  
Expanding ${\psi}(\theta)$ in the basis of plane-wave states 
$\phi_l(\theta) = e^{il\theta}/\sqrt{2\pi}$, we write
\begin{eqnarray}
 \psi(\theta) = \sum_{l} c_{l} \, \phi_{l},
\label{exphinaa}
\end{eqnarray}
where we treat the coefficients $c_l$ as variational parameters.  
As shown in Refs.\,\cite{Leg}, the physical quantities of a system 
described by the Hamiltonian of Eq.\,(\ref{Ham}) are given by the 
phase winding number $J$,
\begin{eqnarray}
   J = [\Omega + 1/2],
\label{jom}
\end{eqnarray}
where $[x]$ is the largest integer that does not exceed $x$, and by the
angular frequency relative to $J$, i.e., $\Omega - J$. Therefore
the dominant term in the expansion of Eq.\,(\ref{exphinaa}) is the one
with $l = J$, where $J$ is given by Eq.\,(\ref{jom}) for a given value
of $\Omega$.  When the density of the gas is uniform, the occupancy 
of all other states is zero.  However when the density is localized 
more states contribute to $\psi$.  Including the $l = J \pm 1$ states
we write,
\begin{eqnarray}
       \psi(\theta) = c_{J-1} \, \phi_{J-1} + c_{J} \, \phi_{J} + 
       c_{J+1} \, \phi_{J+1},
\label{exphaa}
\end{eqnarray}
where in order for $\psi$ to be normalized, $|{c}_{J-1}|^2 + |c_J|^2 + 
|c_{J+1}|^2 =1$.  Close to the phase boundary the dominant terms in 
the order parameter are the ones with $l = J, J \pm 1$, since other 
states (with $l=J \pm 2$, etc.) have relatively higher kinetic energy 
and for this reason we can neglect them.  As a variational method, our 
approach is more reliable close to the phase boundary, while further 
away from it, one needs to include more basis states.

We proceed by expressing the energy per particle ${\epsilon_J}$
in the state of Eq.\,(\ref{exphaa}) in terms of the coefficients $c_l$, 
which are then determined by minimizing ${\epsilon_J}$ with respect to them 
\cite{GK,KMP}. We get that 
\begin{eqnarray}
   {\epsilon_J} = (J - \Omega)^2 |c_{J}|^2  \phantom{XXXXXXXXXXXX}
\nonumber \\
                + (J - 1 - \Omega)^{2} |c_{J-1}|^2 
		 +  (J + 1 - \Omega)^{2} |c_{J+1}|^2
\nonumber \\
   +  ({\gamma} / 2) [|c_{J-1}|^4 + |c_{J}|^4 + |c_{J+1}^4|      
\nonumber \\
 + 4 |c_{J-1}|^2 |c_{J}|^2 + 4 |c_J|^2 |c_{J+1}|^2 + 4 |c_{J-1}|^2 |c_{J+1}|^2  
\nonumber \\
 + 2 c_{J}^2 c_{J-1}^* c_{J+1}^*
   + 2 (c_{J}^2)^* c_{J-1} c_{J+1}],
\label{eneyaa}
\end{eqnarray}
where $\gamma = N U_{0}/(2 \pi)$. If we write $c_l = |c_l| e^{i \theta_l}$, 
there are three phases involved in the problem. One phase is arbitrary, 
the other one is
free (reflecting the rotational invariance of the problem) and we choose it 
here so that the maximum of the density is at $\theta = 0$, and the third
one is chosen so that the energy is minimum. To minimize ${\epsilon_J}$ we
choose $\theta_{J-1} + \theta_{J+1} - 2 \theta_J = 0$ in order to make the 
last two terms in Eq.\,(\ref{eneyaa}) as large as possible \cite{GK}. 
Using the normalization condition we eliminate $c_{J}$ thus getting
\begin{eqnarray}
    {\epsilon_J} - {\gamma} / 2 - (\Omega - J)^2  
      = |c_{J-1}|^2 [1 - 2 (J - \Omega)] 
\nonumber \\
       + |c_{J+1}|^2 [1 + 2 (J - \Omega)]
       \nonumber \\
       + {\gamma}  [- |c_{J-1}|^4 - |c_{J+1}|^4 + 
        |c_{J-1}|^2 + |c_{J+1}|^2 
\nonumber \\
	- |c_{J-1}|^2 |c_{J+1}|^2 +
	2 |c_{J-1}| |c_{J+1}| 
\nonumber \\
	- 2 |c_{J-1}|^3 |c_{J+1}| - 2 |c_{J-1}| |c_{J+1}|^3].
\label{eneybb}
\end{eqnarray}
When the density of the system is uniform, $|c_{J \pm 1}| = 0$, and
the energy per particle is $\epsilon_J = \gamma/2 + (\Omega - J)^2$, 
in agreement with the exact solution of the mean-field approximation 
\cite{Ueda2}.  

From Eq.\,(\ref{eneybb}) we now determine the phase boundary and 
the order parameter.  Minimizing Eq.\,(\ref{eneybb}) we get two 
coupled equations,
\begin{eqnarray}
   1 \mp 2 (J - \Omega) + \gamma [ (\lambda^{\pm 1} + 1) (1 - |c_{J \pm 
   1}|^{2}) - 2 |c_{J \mp 1}|^{2} \nonumber \\
 -  3 |c_{J + 1}| |c_{J - 1}|] = 0,
\label{cenminim}
\end{eqnarray}
where $\lambda = |c_{J+1}|/|c_{J-1}|$.  Close to the phase boundary 
$|c_{J \pm 1}|$ is either zero, or $\ll 1$.  This observation allows 
us to write Eq.\,(\ref{cenminim}) as
\begin{eqnarray}
   1 - 2 (J - \Omega) + \gamma (1 + \lambda) &=& 0
   \nonumber \\
   1 + 2 (J - \Omega) + \gamma (1 + 1/\lambda) &=& 0,
\label{enminim}
\end{eqnarray}
and in order for a solution to exist, $\gamma = -1/2 + 2 (J - 
\Omega)^2$, which is the equation that gives the phase boundary.  Figure 
1 shows this curve and is periodic in $\Omega$, with a period of 
$\Omega = 1$.  For $\gamma$ larger than this critical value the state 
is uniform, while for smaller values it is localized.  Reference 
\cite{Ueda2} has derived the same result by examining the stability of 
the exact mean-field solution.  Furthermore the transition between the 
two phases is of second order, since the order parameter is continuous 
across the phase boundary. Expanding we find that for $\gamma \alt -1/2$
and $\Omega - J \approx 0$,
\begin{eqnarray}
  |c_l|^2 = \frac {1 + 2 \gamma} {7 \gamma} [1 + \alpha_l (\Omega - J)],
\label{ciexp}
\end{eqnarray}
where $\alpha_{J-1} = -18/7$ and $\alpha_{J+1} = 38/7$, while
$|c_J|^2 = 1 - |c_{J-1}|^2 - |c_{J+1}|^2$.

Going even further, we have solved the two coupled Eqs.\,(\ref{cenminim}).
Figure 2 shows the density of the gas $|\psi|^2$ for $\gamma = -0.55$ 
and $-0.45$ in each graph, and for $|\Omega - J| = 0, 1/4$, and 1/2.  
For the smaller value of $\gamma$ the density is more peaked, as one 
expects intuitively.  On the other hand the difference between the two 
states gets smaller as one gets further away from the phase boundary.  
Comparing our variational solution with the exact one (within the 
mean-field approximation) for $\gamma = -0.55$ of Ref.\,\cite{Ueda2}, 
we see that in general the difference is small and it increases as 
$|\Omega - J|$ increases.  This is expected, however, since for higher 
values of $|\Omega - J|$ one is further away from the phase boundary 
and more states have to be included in the expansion of the order 
parameter.  Furthermore, when $|\Omega - J| = 0$ the solution 
coincides with the one of a non-rotating cloud \cite{Ueda,GK}, apart 
from the phase factor $e^{i J \theta}$, while for $|\Omega - J| = 1/2$ 
the order parameter has a node, since at this point the phase jumps by 
$\pi$ and $J$ increases by one unit \cite{Ueda2}.

Let us now turn to the second approach we have developed, namely the 
diagonalization of the Hamiltonian $\hat H$.  Following the same steps 
as before, we expand the field operator $\hat{\psi}$, and for the same 
reasons as before,
\begin{eqnarray}
    \hat{\psi}(\theta) = \hat{c}_{J-1} \, \phi_{J-1} + \hat{c}_J \, \phi_{J}  
   + \hat{c}_{J+1} \, \phi_{J+1},
\label{exph}
\end{eqnarray}
where in this case $\hat{c}_l$ is the annihilation operator of an atom 
with angular momentum $l$. When the density of the gas is uniform, 
the states $l = J \pm 1$ are equally occupied and their occupancy goes to zero 
in the $N \to \infty$ limit.  In addition the occupancy of other 
states ($l=J \pm 2$, etc.) goes more rapidly to zero (as $N \to 
\infty$), since these states have relatively higher kinetic energy and 
for this reason we neglect them.  Writing the Hamiltonian (\ref{Ham}) 
in second-quantized form,
\begin{eqnarray} 
   \hat{H}= \sum_{l} (l-\Omega)^2 \hat{c}_l^{\dagger} 
  \hat{c}_l + \frac{U_0}{4\pi} \sum_{klmn} \hat{c}^{\dagger}_k 
 \hat{c}_l^{\dagger} \hat{c}_m \hat{c}_n \, \delta_{m+n-k-l},
\label{Haml}
\end{eqnarray}
we notice that it can be diagonalized with a Bogoliubov transformation 
in the subspace of states with $l=J, J \pm 1$, as we show below \cite{KMP}.  
We work with the basis vectors
\begin{equation}
  |m \rangle = |N_{J-1},N_J,N_{J+1} \rangle \equiv |(J-1)^m, J^{N-2m}, (J+1)^m 
\rangle,
\label{mu}
\end{equation}
where $N_l$ is the occupancy of the state with angular momentum $l$. 
The states $| m \rangle$ have been constructed to have $N$ atoms, $\sum_l N_l = 
N$, $N J$ units of angular momentum, $\sum_l l \, N_l= N J$, and an equal 
occupancy of the $l = J \pm 1$ states (i.e., $m$ atoms in each of them).  The 
diagonal matrix elements are,
\begin{eqnarray}
    \langle m | \hat{H} | m \rangle = m [(J - 1 - \Omega)^{2} + (J +1 - 
    \Omega)^{2}] 
\nonumber \\
+ (N - 2m) (J - \Omega)^{2} + \frac {U_0} {4 \pi} 
    [(N-2m)(N-2m-1) \nonumber \\
 + 2 m (m-1) + 8 m (N-2m) + 4 m^2].
\end{eqnarray}
When the system is on the side of the ``uniform state" (in the limit 
$N \to \infty$) then $m$ is of order unity, which allows us to neglect 
the terms of order $m^2$ compared to those of order $N^{2}$ and $N m$ 
(on the contrary, $m$ becomes of order $N$ when a localized state forms),
\begin{eqnarray}
    \langle m | \hat{H} | m \rangle \approx N (J - \Omega)^{2} 
   + \gamma (N-1)/2 + 2 m (1 + \gamma).
\label{meappr1}
\end{eqnarray}

We also get for the off-diagonal matrix elements,
\begin{eqnarray}
\langle m | \hat{H} | m+1 \rangle &=&
     \frac {U_0} {4 \pi} 2 \sqrt{(N-2m) (N-2m-1) (m+1)^2} 
\nonumber \\
     &\approx& \gamma (m+1)
\label{meappr}
\end{eqnarray}
in the limit $m \ll N$. From Eqs.\,(\ref{meappr1}) and 
(\ref{meappr}) $\hat{H}$ can be written as
\begin{eqnarray}
\hat{H} - \gamma (N-1)/2 - N (J - \Omega)^{2} = 
\nonumber \\
 (1 + \gamma) (\hat{c}_{J-1}^{\dagger} \hat{c}_{J-1} +
		\hat{c}_{J+1}^{\dagger} \hat{c}_{J+1})  
\nonumber \\
   + \gamma (\hat{c}_{J-1}^{\dagger} \hat{c}_{J+1}^{\dagger} +
		\hat{c}_{J-1} \hat{c}_{J+1}).
\label{hamapl}
\end{eqnarray}
The above Hamiltonian can be diagonalized with use of a Bogoliubov
transformation \cite{Bog}, i.e., introducing the operators
\begin{eqnarray}
 \hat{b} = \lambda_1 \hat{c}_{J-1}^{\dagger} + \lambda_2 \hat{c}_{J+1}
\,\,\, {\rm and} \,\,\,
 \hat{d} = \lambda_2 \hat{c}_{J-1} + \lambda_1 \hat{c}_{J+1}^{\dagger}.
\end{eqnarray}
Following the usual tricks, we get that the eigenvalues of $\hat{H}$ 
are given by
\begin{eqnarray}
{\cal E}_{n_b,n_d} - \gamma (N-1)/2 - N (J - \Omega)^{2} = 
\nonumber \\
   - (\gamma +1) + 
\sqrt{1 + 2 \gamma} (1 + n_b + n_d),
\label{hamapldd}
\end{eqnarray}
where $n_b, n_d$ are the eigenvalues of the number operators  
$\hat{b}^{\dagger} \hat{b}$ and $\hat{d}^{\dagger} \hat{d}$ 
respectively.  To get the above solution, $\gamma$ has to be larger 
than $-1/2$.  For $N \to \infty$, one can identify the term ${\cal 
E}_{0,0}/N = \gamma/2 + (J - \Omega)^{2}$ as the total energy of the 
condensate with uniform density, in agreement with our variational 
approach and with the exact solution of the mean-field theory, 
Ref.\,\cite{Ueda2}.

Furthermore, one can see that for the states $|m \rangle$, 
$(\hat{b}^{\dagger} \hat{b} - \hat{d}^{\dagger} \hat{d}) |m \rangle = 0,$
which implies that $n_b = n_d = 0,1,2, \dots$  Therefore, the first 
excited state of the system is (measured from the lowest energy) 
$2 \sqrt{1 + 2 \gamma}$ as Eq.\,(\ref{hamapldd}) implies, if $n_{b} = 
n_{d} = 1$.  This result is in agreement with Ref.\,\cite{Ueda2}.

Up to now the only condition we have found in order for a solution 
to exist is that $\gamma > -1/2$.  Turning to the stability of our solution 
against small perturbations, we consider two states where one atom with 
angular momentum $l=J-1$ has been transferred to the state with 
$l=J+1$, or vice versa,
\begin{eqnarray}
   |m^{\pm} \rangle = |(J-1)^{m \mp 1}, J^{N-2m}, (J+1)^{m \pm 1} \rangle.
\label{muexc}
\end{eqnarray}
The states $|m^{\pm} \rangle$ have $N$ atoms, and they carry an angular 
momentum that differs by two units compared to that of the state $|m\rangle$, 
$\sum_{l} l N_{l} = N J \pm 2$.  One can follow the same steps as 
before and get that the energy is
\begin{eqnarray}
  {\cal E}^{\pm}_{n_b,n_d} - \gamma (N-1)/2 - N (J - \Omega)^{2} \mp 4 
  (J - \Omega) = \nonumber \\
 - (\gamma +1) + \sqrt{1 + 2 \gamma} (1 + n_b + n_d).
\label{hamaplddexc}
\end{eqnarray}
In this case $(\hat{b}^{\dagger} \hat{b} - \hat{d}^{\dagger} \hat{d})
|m^{\pm} \rangle = \pm 2 |m^{\pm} \rangle$ 
and thus the lowest energy of the system is that with $n_{b} = 2,  
n_{d} = 0$, or vice versa.  In order for our solution to be stable, 
the lowest energy of ${\cal E}^{\pm}_{n_b,n_d}$ has to be higher than 
that of ${\cal E}_{n_b,n_d}$, or
\begin{equation}
  \sqrt{1 + 2 \gamma} \ge 2 |\Omega - J|,
\label{condstab}
\end{equation}
which can also be written as $\gamma > -1/2 + 2 (J - \Omega)^2$, in 
agreement with the expression we derived earlier within our variational
scheme. If Eq.\,(\ref{condstab}) is not satisfied, even an infinitesimally 
small anisotropy in the trapping potential suffices to convert the density 
of the system from uniform to localized.  

Finally we calculate the depletion of the condensate $\Delta N$,
\begin{equation}
  \Delta N = \langle \hat{c}_{J-1}^{\dagger} \hat{c}_{J-1} \rangle + 
             \langle \hat{c}_{J+1}^{\dagger} \hat{c}_{J+1} \rangle 
           = \frac {\gamma + 1} {\sqrt{1 + 2 \gamma}} - 1,
\label{fluct}
\end{equation}
which is singular only when $J = \Omega$.  More precisely, close to 
the phase boundary and for small values of $|J - \Omega|$, $\Delta N 
\approx 1/(4 |J - \Omega|)$, but it is finite otherwise.  These 
results are also in agreement with those of the mean-field approach
\cite{Ueda2}.

To summarize, we examined a one-dimensional Bose-Einstein condensate 
of atoms confined in a rotating torus, using two independent methods, 
namely a variational mean-field approach, as well as an exact  
diagonalization of the Hamiltonian.  Based on these models, we 
examined the two phases which appear of uniform and localized density, 
and we identified the phase boundary which separates them, as the 
strength of the interaction and the frequency of rotation vary.  Our 
results are in agreement with those of the exact mean-field approach 
and with numerical diagonalization of the Hamiltonian of the system 
\cite{Ueda2}.

\vskip1.0pc
\noindent The author is grateful to E. Lundh and N. Papanicolaou for fruitful
discussions and to N. Kylafis for his assistance. The author also wishes
to thank the Physics Department of the Univ. of Crete for its hospitality.

\begin{figure}
\begin{center}
\epsfig{file=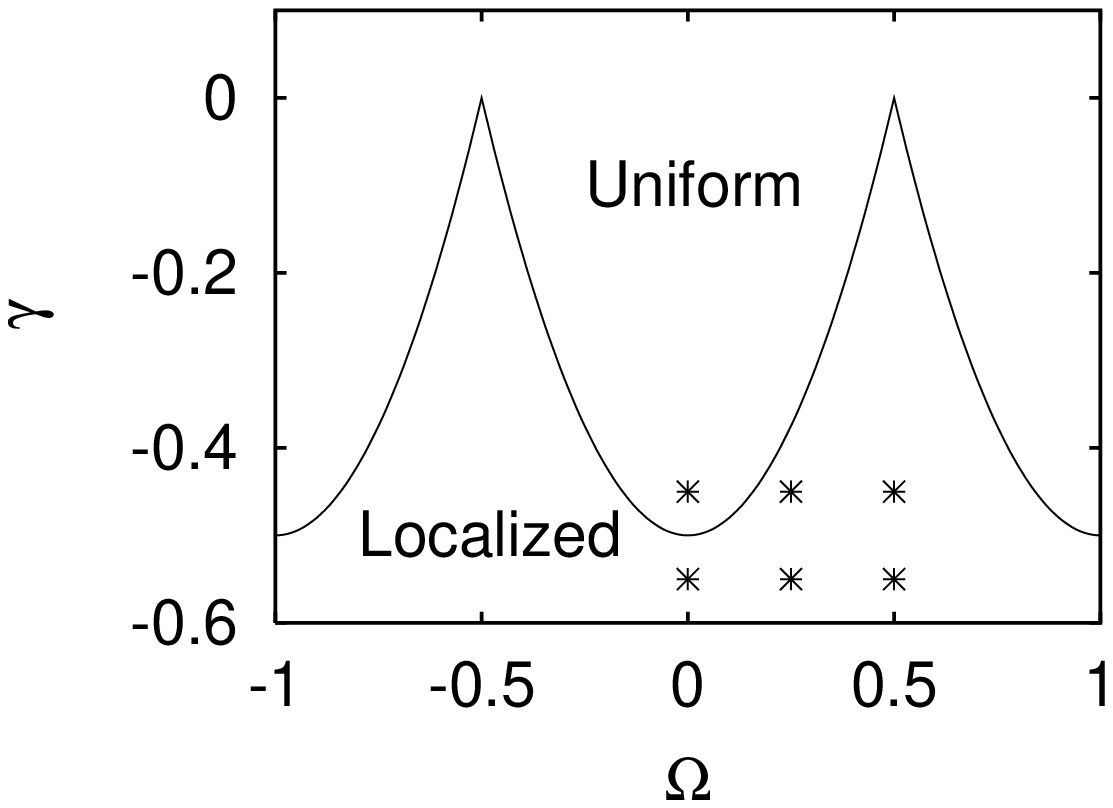,width=8.0cm,height=5.0cm,angle=0} 
\vskip0.5pc
\begin{caption}
{The phase boundary between uniform and localized 
density, $\gamma = -1/2 + 2 (J - \Omega)^{2}$.  The curve is periodic 
in $\Omega$ with a period equal to 1.  The stars denote the six cases 
considered in Fig.\,2.}
\end{caption}
\end{center}
\label{FIG1}
\end{figure}

\begin{figure}
\begin{center}
\epsfig{file=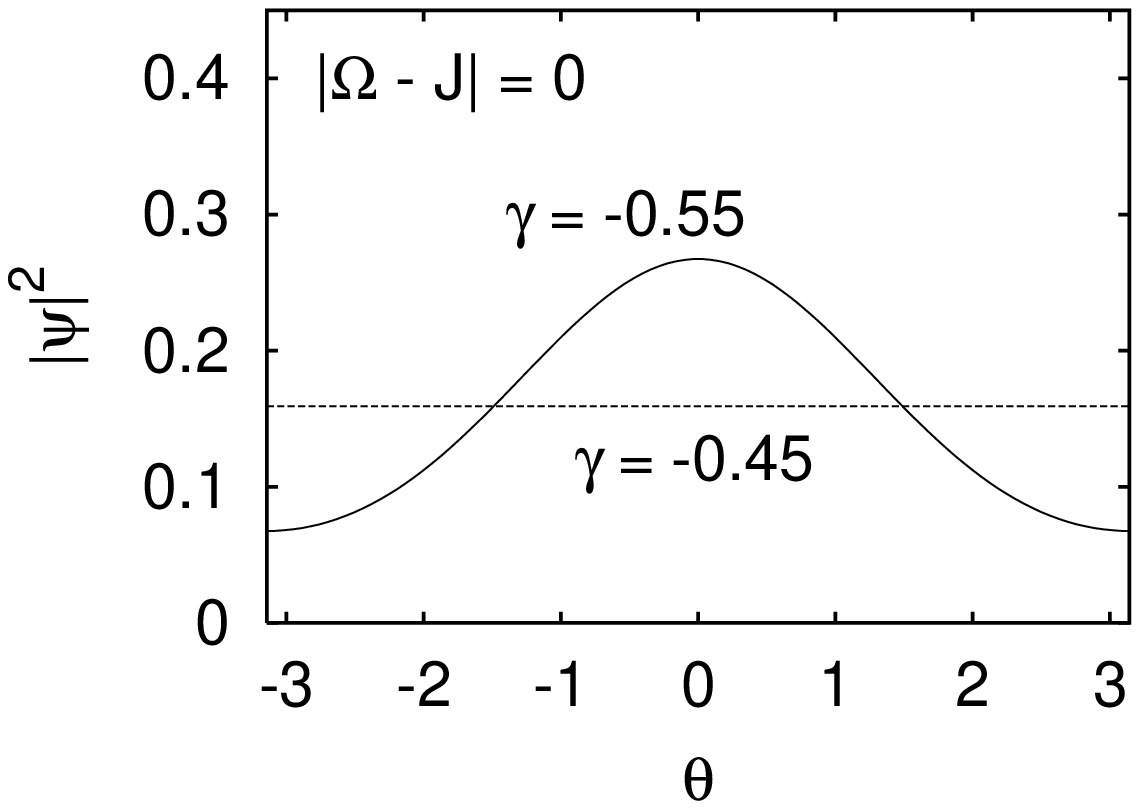,width=8.0cm,height=4.4cm,angle=0}
\epsfig{file=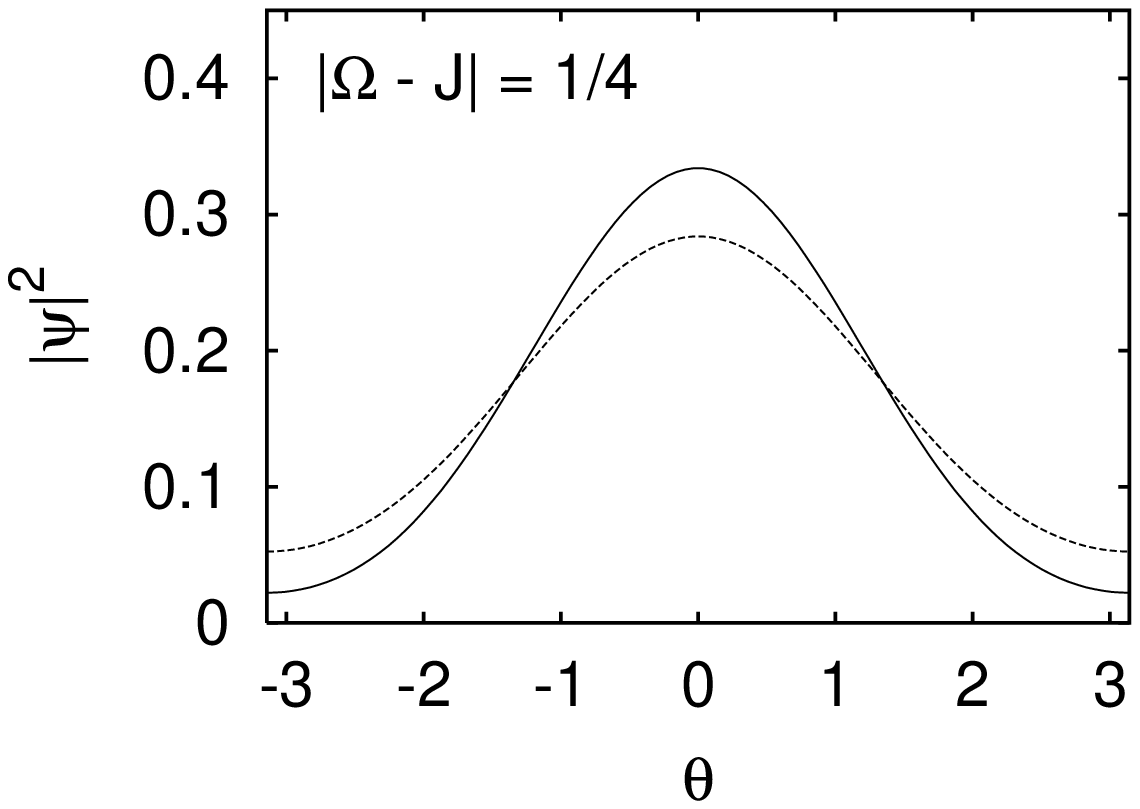,width=8.0cm,height=4.4cm,angle=0}
\epsfig{file=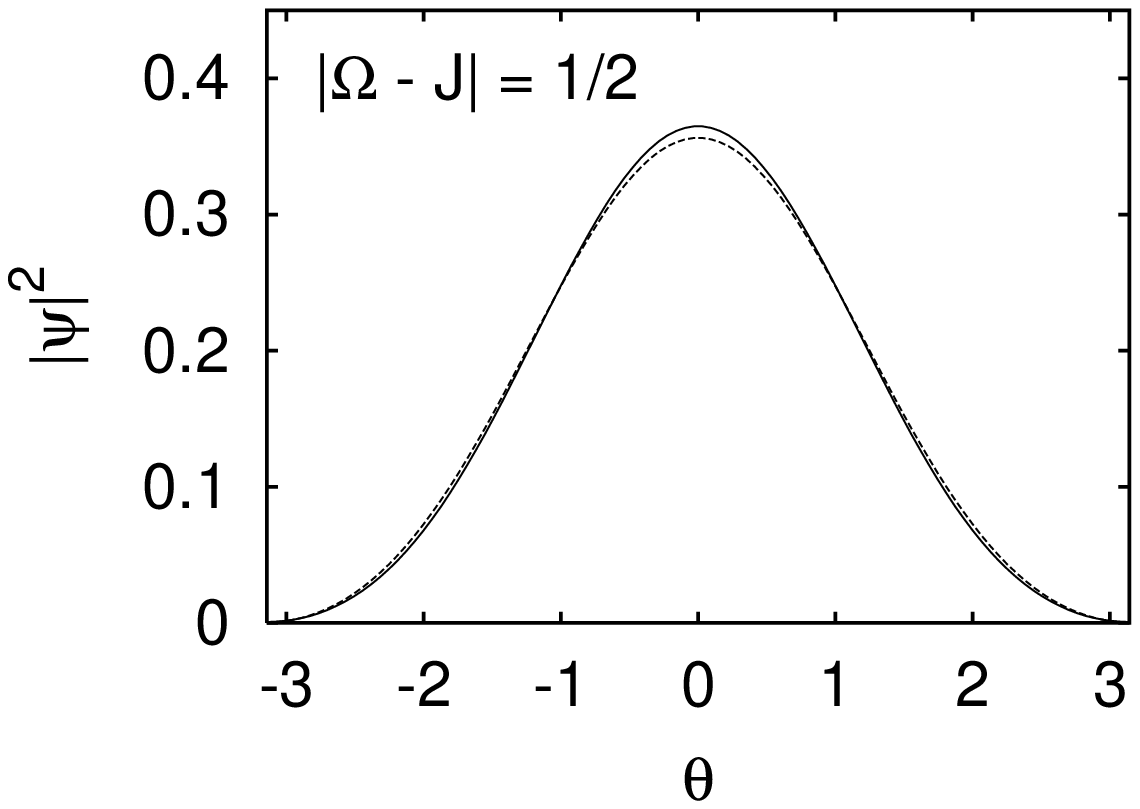,width=8.0cm,height=4.4cm,angle=0}
\vskip0.5pc
\begin{caption}
{The density of the trial order parameter $|\psi|^2$, Eq.\,(4), 
calculated within the mean-field approximation,  for $|\Omega - J| = 0$
(top) $|\Omega - J| = 1/4$ (middle), and $|\Omega - J| = 1/2$ (bottom) 
with $\gamma = -0.55$ in the solid curves and $-0.45$ in the dashed curves.}  
\end{caption}
\end{center}
\label{FIG2}
\end{figure}


\begin{references}

\bibitem{Kettnew} This limit was first reached experimentally in a 
Bose-Einstein condensate of $^{23}$Na atoms; see: A.~G\"orlitz, 
J.~M.~Vogels, A.~E.~Leanhardt, C.~Raman, T.~L.~Gustavson, 
J.~R.~Abo-Shaeer, A.~P.~Chikkatur, S.~Gupta, S.~Inouye, 
T.~P.~Rosenband, D.~E.~Pritchard, and W.~Ketterle, Phys.  Rev.  Lett.  
{\bf 87}, 130402 (2001).

\bibitem{sol1} K.~E.~Strecker, G.~B.~Partridge, A.~G.~Truscott, and
R.~G.~Hulet, Nature (London) {\bf 417}, 150 (2002).

\bibitem{sol2} L.~Khaykovich, F.~Schreck, G.~Ferrari, T.~Bourdel,
J.~Cubizolles, L.~D.~Carr, Y.~Castin, and C.~Salomon, Science {\bf 296}, 1290
(2002).

\bibitem{Ket} S.~Inouye, M.~R.~Andrews, J.~Stenger, H.~-J.~Miesner,
D.~M.~Stamper-Kurn, and W.~Ketterle, Nature (London) {\bf 392}, 151 (1998).

\bibitem{refs} L.~D. Carr and Y.~Castin, Phys. Rev. A {\bf 66}, 063602 (2002);
U.~Al~Khawaja, H.~T.~C. Stoof, R.~G.~Hulet, K.~E.~Strecker, and
G.~B.~Partridge, Phys. Rev. Lett. {\bf 89}, 200404 (2002); L.~Salasnich,
A.~Parola, and L.~Reatto, Phys. Rev. A {\bf 66}, 043603 (2002).

\bibitem{Ueda} R.~Kanamoto, H.~Saito, and M.~Ueda, Phys. Rev. A {\bf 67},
013608 (2003).

\bibitem{GK} G.~M.~Kavoulakis, Phys.  Rev.  A {\bf 67}, 011601(R) (2003).

\bibitem{Ueda2} R.~Kanamoto, H.~Saito, and M.~Ueda, e-print cond-mat/0305307.

\bibitem{Emil} E.~Lundh, A.~Collin and K.-A.~Suominen, e-print cond-mat/0304186.

\bibitem{pr} E.~Lundh, private communication.

\bibitem{Leg} A.~J.~Leggett, {\it Low Temperature Physics}, edited
by M.~J.~R.~Hoch and R.~H.~Lemmer (Springer Verlag, Berlin); Physica
Fennica {\bf 8}, 125 (1973).

\bibitem{KMP} A similar problem has been attacked in the 
context of weakly-interacting Bose-Einstein condensates under 
rotation in: G.~M.~Kavoulakis, B.~Mottelson, and C.~J.~Pethick, Phys.  
Rev.  A {\bf 62}, 063605 (2000).

\bibitem{Bog} N.~N.~Bogoliubov, J. Phys. (Moscow) {\bf 11}, 23 (1947).

\end{references}
\end{document}